\documentclass[preprint,prc,aps,showpacs,showkeys,groupedaddress,floatfix]{revtex4-1}
\usepackage{epsfig}
\usepackage{dcolumn}
\usepackage{bm}
\usepackage[latin5]{inputenc}
\usepackage{graphics}
\usepackage{graphicx}
\usepackage{epsfig}
\usepackage{amssymb}
\usepackage{amsmath}
\begin{document}
\title{Corrected Analytical Solution of the Generalized Woods-Saxon Potential for Arbitrary $\ell$ States}
\author{O. Bayrak}
\affiliation{Department of Physics, Akdeniz University, 07058 Antalya,
Turkey}
\author{E. Aciksoz}
\affiliation{Department of Physics, Akdeniz University, 07058 Antalya,
Turkey}
\published{ in Phys. Scr. \textbf{90 }(2015) 015302}
\begin{abstract}
The bound state solution of the radial Schr\"{o}dinger equation with
the generalized Woods-Saxon potential is carefully examined by using
the Pekeris approximation for arbitrary $\ell$ states. The energy
eigenvalues and the corresponding eigenfunctions are analytically
obtained for different $n$ and $\ell$ quantum numbers. The obtained
closed forms are applied to calculate the single particle energy
levels of neutron orbiting around $^{56}$Fe nucleus in order to
check consistency between the analytical and Gamow code results. The
analytical results are in good agreement with the results obtained
by Gamow code for $\ell=0$.
\end{abstract}

\maketitle

\section{Introduction}
\begin{figure}
\includegraphics[width=\textwidth]{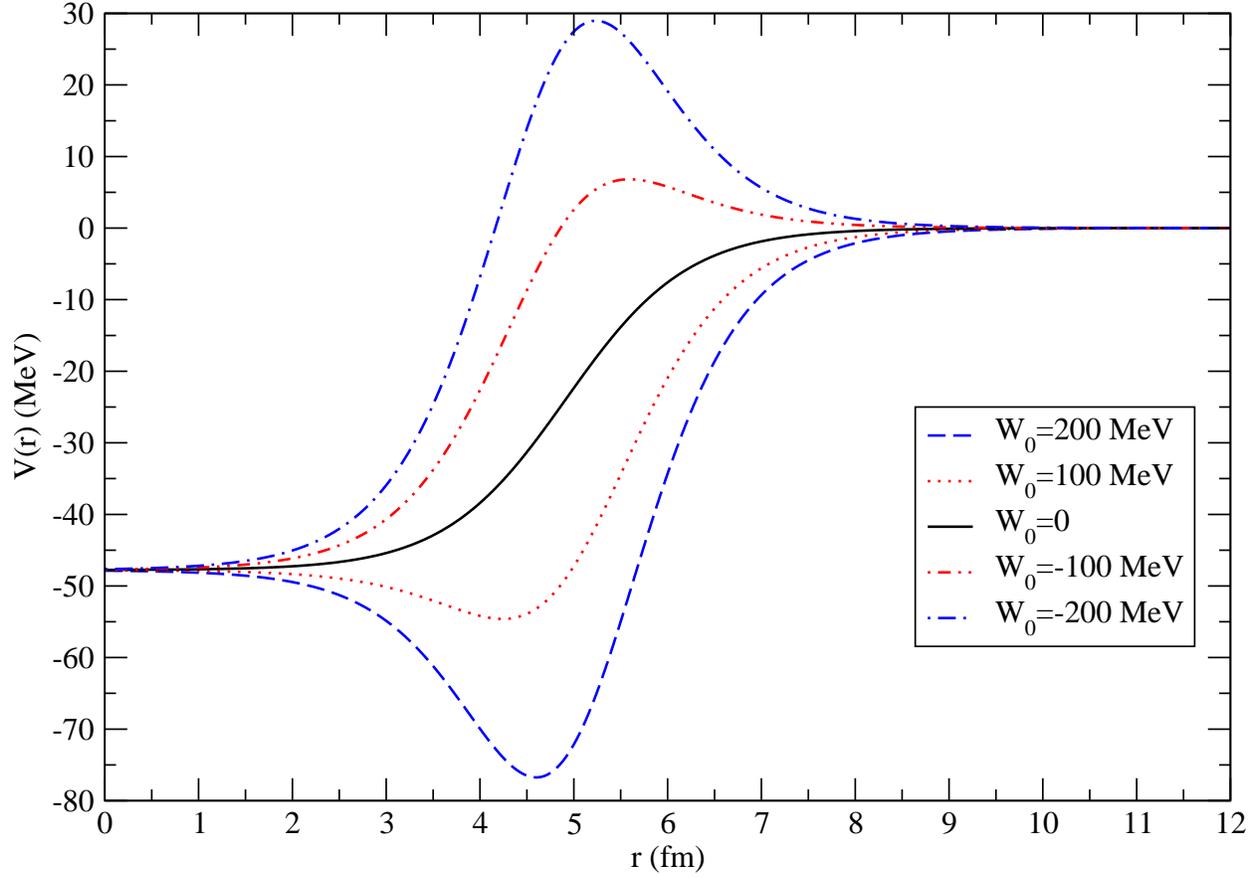}
\caption{Variation of the generalized Woods-Saxon potential as a
function of the internuclear distance and several $W_{0}$ values for
$V_{0}$$=$40.5+0.13A MeV, R$=$4.9162 fm and a$=$0.6 fm. }
\label{potplot}
\end{figure}

The Woods-Saxon potential was firstly proposed by R.D. Woods and
D.S. Saxon in order to explain the elastic scattering of 20 MeV
protons by medium and heavy nuclei approximately sixty years
ago\cite{Wood}. Thereafter the Woods-Saxon potential has taken a
great deal of interest over the years and has been one of the most
useful model to determine the single particle energy levels of
nuclei\cite{bound,Gomez,Massen} and the nucleus-nucleus
interactions\cite{bound-scat,scattering,Khoa}. The modified version
of the Woods-Saxon potential consists of the Woods-Saxon and its
derivative called the Woods-Saxon surface potential and is given
by\cite{EckartWS,modifiedWs,berkdemir}, 
\begin{equation}\label{ws}
  V(r)=-\frac{V_{0}}{1+e^{(\frac{r-R}{a})}}-\frac{W_{0}
  e^{(\frac{r-R}{a})}}{(1+e^{(\frac{r-R}{a})})^{2}},
\end{equation}
where $V_{0}$ and $W_{0}$ represent the depths of the potential
well. $R$ and $a$ are the radius of the potential and the width of
the surface diffuseness, respectively. In Fig.\ref{potplot}, a form
of the generalized Woods-Saxon potential (GWS) versus the
internuclear distance is shown for the given potential parameters
with several $W_{0}$ values. The surface term in the generalized
Woods-Saxon potential induces an extra potential pocket, especially,
at the surface region of the potential, and this pocket is very
important in order to explain the elastic scattering of some nuclear
reactions\cite{bound-scat,scattering}. Moreover, the Woods-Saxon
surface potential induces a potential barrier for $W_{0}<0$ so that
it could be used in explaining of the resonant states (quasi-bound
states) in nuclei. There are some special cases of the generalized
Woods-Saxon potential: The GWS potential is reduced to the standard
Woods-Saxon form for $W_{0}=0$ and the square well potential for
$W_{0}=0$ and $a\rightarrow0$\cite{fluge}. Furthermore the GWS
potential is reduced to the Rosen-Morse potential\cite{susy} for
$R=0$\cite{editor}.

The relativistic treatment of a Dirac particle in the Woods-Saxon
potential field is examined in three dimensions for
$\ell=0$\cite{normaldirac}. Moreover, the transmission coefficient
and bound state solutions of one dimensional Woods-Saxon potential
are analytically studied\cite{Kennedy}. The Klein-Gordon equation in
the presence of a spatially one-dimensional Woods-Saxon potential is
also examined. The scattering state solutions are obtained in terms
of hypergeometric functions and the condition for the existence of
transmission resonances is derived\cite{KG}. Furthermore s-wave
solution of the Dirac equation for a particle moving in the
spherically symmetric Woods-Saxon potential under the conditions of
the exact spin and the pseudospin symmetry limit is
examined\cite{spin} and is discussed in
Refs.\cite{spincomment,spinreply}.

It is known that the exact analytical solutions of the wave
equations (Schr\"{o}dinger, Dirac, etc.) are very important since a
closed form of the wave function is more convenient than the wave
function obtained by numerical calculation in explaining the
behavior of the system under consideration. Unfortunately, there are
few potentials such as harmonic oscillator, Coulomb and Kratzer
potentials \emph{etc.}\cite{fluge} which have the exact analytical
solution with centrifugal term. In literature, there are some effort
about obtaining the approximate analytic solutions of the wave
equations in terms of the $\ell\neq0$ case: The most widely used
approximation is introduced by Pekeris\cite{Pekeris} for the
exponential-type potential so that this approximation is based on
the expansion of the centrifugal barrier in a series of exponentials
depending on the internuclear distance up to second order. Greene
and Aldrich\cite{Greene} proposed another approximation for the
centrifugal term $1/r^{2} \approx \delta^{2} e^{\delta
r}/(1-e^{\delta r})^{2}$. However, this approximation is valid only
for small values of the screening parameter $\delta$\cite{BayrakG}.

Recently, the GWS potential has been examined by using the
Nikiforov-Uvarov (NU) method\cite{Nifikorov} for $\ell=0$
state\cite{berkdemir}. However, in the paper authors have obtained
the energy eigenvalue equation as $R$ independent due to $r-R=r$
transformation so that the potential is reduced to the Rosen-Morse
potential\cite{editor}. It has been noted that the analytical and
the numerical results are inconsistent for the GWS potential with
$\ell=0$ state\cite{editor}. Similar results with
Ref.\cite{berkdemir} can be found in
Refs.\cite{ikot,gonul,badalovWS,badalovKGWS} for the relativistic or
non-relativistic wave equations. In Ref.\cite{ikot}, the approximate
analytical solution of the Schr\"{o}dinger equation for the standard
Woods-Saxon potential is obtained for any $\ell$ state by using the
NU method. In Ref.\cite{gonul}, the GWS potential is examined for
the Klein-Gordon and Schr\"{o}dinger equations. In
Ref.\cite{badalovWS,badalovKGWS} the Woods-Saxon potential has been
analyzed for both the radial Schr\"{o}dinger and Klein-Gordon
equations by using the Pekeris approximation. The authors in
Refs.\cite{badalovWS,badalovKGWS} have used
$z(r)$$=$$\frac{1}{1+e^{\frac{r-R}{a}}}$ transformation and have
obtained $R$ dependent eigenvalue equation by using the
Nikiforov-Uvarov method\cite{badalovWS,badalovKGWS}. However, we
have realized that the Nikiforov-Uvarov method can not take into
account the boundary condition correctly since the Woods-Saxon
potential has different character close to r$=$R. Therefore, in this
article, we have carefully examined the radial Schr\"{o}dinger
equation with the generalized Woods-Saxon potential by using the
Pekeris approximation in terms of the correct boundary conditions.
In the next section, we present the calculation procedure. Then, in
section \ref{conclude} is devoted to the summary and conclusion.

\section{The Energy Eigenvalues and Eigenfunctions}
The generalized Woods-Saxon potential or special forms of it are
very useful in order to describe the interactions between the
systems, especially, in nuclear physics. In order to explain the
single particle energy levels or elastic scattering of nuclei, the
Woods-Saxon potential is generally used. Since the interactions
usually occur at the surface region of nuclei for both bound and
continuum states, the form of the potential at the surface is
crucially important. Therefore the surface term in Eq.(\ref{ws})
would be a very convenient model in order to calculate the single
particle energy levels of nuclei. When we consider two-body system
with the reduced mass $\mu$ moving under the generalized Woods-Saxon
potential, the effective potential is, 
\begin{equation}\label{veff}
    V_{eff}(r)=V(r)+V_{\ell}(r)=-\frac{V_{0}}{1+e^{\frac{r-R}{a}}}-\frac{W_{0} e^{\frac{r-R}{a}}}{(1+e^{\frac{r-R}{a}})^{2}}+ \frac{\ell(\ell+1)\hbar^{2}}{2\mu
    r^{2}},
\end{equation}
where $\mu=\frac{m_nm_A}{m_n+m_A}$. $m_n$ and $m_A$ are the atomic
mass of the neutron and the core nucleus respectively. There is no
analytical solution of Eq.(\ref{veff}) due to polynomial form of the
centrifugal barrier term. Therefore, we have to use an approximation
for the centrifugal term similar to other authors\cite{badalovWS}.
In literature there are few approximation
procedure\cite{Pekeris,Greene}. One of them is the Pekeris
approximation\cite{Pekeris} based on an expansion of the centrifugal
barrier depending on the internuclear separation up to second
order\cite{badalovWS}.

The quasi-analytical solution of the effective potential in
Eq.(\ref{veff}) with the Pekeris approximation \cite{Pekeris} can be
obtained within the framework of the Nikiforov-Uvarov (NU) or
asymptotic iteration (AIM) methods  as follows, 
\begin{equation}\label{nueigen}
   n_{r}(n_{r}+1)-\beta^2-\gamma_2^2+(1+2n_{r})\varepsilon+2\varepsilon^2+(1+2n_{r}+2\varepsilon)\sqrt{\varepsilon^2+\gamma_1^2-\beta^2}=0, \,\,
   n_{r}=0,1,2....
\end{equation}
with the following definitions: 
\begin{eqnarray}\label{ansatz}
-\varepsilon^{2}=\frac{2\mu a^{2}(E-\delta C_0)}{\hbar^{2}}, \quad
\beta^{2}=\frac{2\mu a^{2}(V_{0}-\delta C_1)}{\hbar^{2}}, \quad
\gamma_{1}^{2}=\frac{2\mu a^{2}\delta C_2}{\hbar^{2}}, \quad
\gamma_{2}^{2}=\frac{2\mu a^{2}W_0}{\hbar^{2}},\\ \quad
C_0=1-\frac{4}{\alpha}+\frac{12}{\alpha^{2}}, \quad
C_1=\frac{8}{\alpha}-\frac{48}{\alpha^{2}}, \quad
C_2=\frac{48}{\alpha^{2}}, \quad
\alpha=\frac{R}{a}, \quad
\delta=\frac{\ell(\ell+1)\hbar^{2}}{2\mu R^{2}}.\nonumber
\end{eqnarray}
In Ref.\cite{editor}, the consistency of the analytical results of
Ref.\cite{berkdemir} with the Gamow code\cite{gamow} has been
checked by calculating the single particle energy levels of the
neutron moving around the $^{56}$Fe nucleus for the given potential
parameters. As can be seen in Ref.\cite{editor}, the results are
inconsistent with the numerical calculations. We have also confirmed
that there are inconsistencies between Eq.(\ref{nueigen}) and the
Gamow code results for $\ell=0$. In Eq.(\ref{nueigen}), if one uses
$W_0=0$, the results of Ref.\cite{badalovWS} can be obtained for
arbitrary $\ell$ states. Furthermore if we take $\ell=0$ in
Eq.(\ref{nueigen}), we get the results of Ref.\cite{berkdemir}.
Therefore we should say that Eq.(\ref{nueigen}) determined by the NU
method is incorrect. If we used any analytical solution methods such
as the asymptotic iteration method (AIM)\cite{saha}, the
supersymmetry (SUSY)\cite{susy}, \emph{etc.} to solve the
corresponding equations with the generalized Woods-Saxon potential,
we would find same results in Eq.(\ref{nueigen}). In literature,
there are similar calculations for the analytical solution of the
generalized Woods-Saxon potential with the relativistic or
non-relativistic wave equations by using the analytical
methods\cite{nuothers}. The origin of the problem is due to the
boundary conditions so that the analytical methods cannot take into
account them correctly since the Woods-Saxon potential has different
characteristic neighborhood r$=$R. Therefore we carefully have to
examine the boundary conditions. In order to get the asymptotic
behavior of the wave function $u_{n\ell}(z)$, we can use the
Nikiforov-Uvarov method\cite{Nifikorov} and easily get
$\phi(z)=z^\varepsilon
(1-z)^{\sqrt{\varepsilon^2-\beta^2+\gamma_1^2}}$. As a result, the
asymptotic solution of the wave function is, 
\begin{equation}\label{aswave}
    u_{n\ell}(z)=z^\varepsilon(1-z)^\eta f_{n\ell}(z),
\end{equation}
where $z$$=$$\frac{1}{1+e^{\frac{r-R}{a}}}$ and
$\eta^2=\varepsilon^2-\beta^2+\gamma_1^2$. The wave function in
Eq.(\ref{aswave}) satisfies the boundary conditions, \emph{i.e.},
$u_{n\ell}(r\rightarrow0,z\rightarrow1)\rightarrow0$ and
$u_{n\ell}(r\rightarrow\infty,z\rightarrow0)\rightarrow0$. The
Schr\"{o}dinger equation becomes for Eq.(\ref{aswave}), 
\begin{equation}\label{radyalf}
 z(1-z)\frac{d^{2}f_{n\ell}(z)}{dz^{2}}+[1+2\varepsilon-(2+2\varepsilon+2\eta)z]\frac{df_{n\ell}(z)}{dz}
 -[-(\gamma_1^2+\gamma_2^2)+\varepsilon+\eta+(\varepsilon+\eta)^2]f_{n\ell}(z)=0.
\end{equation}
It is known that the hypergeometric equation\cite{Abramowitz} is
defined as
\begin{equation}\label{hyper}
 z(1-z)\frac{d^{2}w(z)}{dz^{2}}+[c-(a+b+1)z]\frac{dw(z)}{dz}-abw(z)=0,
\end{equation}
and one of the solutions is
$w(z)={_{2}}F_{1}(a,b;c;z)$\cite{Abramowitz}. In order to get
$a,b,c$ parameters we compare Eq.(\ref{radyalf}) with
Eq.(\ref{hyper}) and find,
\begin{eqnarray}\label{hyperabc}
 a&=&\frac{1}{2}(1\mp\sqrt{1+4\gamma_1^2+4\gamma_2^2}+2\varepsilon+2\eta), \nonumber \\
 b&=&\frac{1}{2}(1\pm\sqrt{1+4\gamma_1^2+4\gamma_2^2}+2\varepsilon+2\eta), \nonumber \\
 c&=&1+2\varepsilon.
\end{eqnarray}
Consequently we have $u_{n\ell}(z)=z^\varepsilon(1-z)^\eta
{_{2}}F_{1}(a,b;c;z)$. To study in the vicinity of $z=1$, we use the
relation\cite{fluge}, 
\begin{eqnarray}\label{f21}
{_{2}}F_{1}(a,b;c;y)&=&\frac{\Gamma(c)\Gamma(c-a-b)}{\Gamma(c-a)\Gamma(c-b)}{_{2}}F_{1}(a,b;a+b-c+1;1-y)\nonumber \\
&+&\frac{\Gamma(c)\Gamma(a+b-c)}{\Gamma(a)\Gamma(b)}(1-y)^{c-a-b}{_{2}}F_{1}(c-a,c-b;c-a-b+1;1-y).
\end{eqnarray}
${_{2}}F_{1}(a,b;c;0)$ is equal 1\cite{Abramowitz}. Therefore, by
using Eq.(\ref{f21}) and the boundary condition
$u_{n\ell}(r\rightarrow0,z\rightarrow1)=0$, we get
\begin{eqnarray}\label{eigen1}
\frac{\Gamma(a+b-c)}{\Gamma(c-a-b)}\frac{\Gamma(c-a)}{\Gamma(b)}\frac{\Gamma(c-b)}{\Gamma(a)}(1+e^{R/a})^{2\eta}=-1,
\end{eqnarray}
where $\eta=i\lambda$ and
$\lambda=\sqrt{\beta^2-\gamma_1^2-\varepsilon^2}$. Evaluating
Eq.(\ref{eigen1}) for the given a,b,c parameters in
Eq.(\ref{hyperabc}), we obtain
\begin{eqnarray}\label{eigen2}
\frac{\Gamma(2i\lambda)}{\Gamma(-2i\lambda)}
\frac{\Gamma(\frac{1}{2}+\frac{1}{2}\sqrt{1+4\gamma_1^2+4\gamma_2^2}+\varepsilon-i\eta)}
{\Gamma(\frac{1}{2}-\frac{1}{2}\sqrt{1+4\gamma_1^2+4\gamma_2^2}+\varepsilon+i\eta)}
\frac{\Gamma(\frac{1}{2}-\frac{1}{2}\sqrt{1+4\gamma_1^2+4\gamma_2^2}+\varepsilon-i\eta)}
{\Gamma(\frac{1}{2}+\frac{1}{2}\sqrt{1+4\gamma_1^2+4\gamma_2^2}+\varepsilon+i\eta)}\nonumber\\
\times(1+e^{R/a})^{2\eta}=-1.
\end{eqnarray}
In Eq.(\ref{eigen2}), $e^{R/a}\gg1$ for a given realistic
parameters. Therefore we can use an approximation $1+e^{R/a}\approx
e^{R/a}$ with small errors in the eigenvalues. We can easily get the
following equation by using
$e^{-2iarg\Gamma(z)}=\frac{\Gamma^{*}(z)}{\Gamma(z)}$ relation, 
\begin{eqnarray}\label{eigen3}
e^{2i\left[arg\Gamma(2i\lambda)-arg\Gamma(\frac{1}{2}+\frac{1}{2}\sqrt{1+4\gamma_1^2+4\gamma_2^2}+\varepsilon+i\eta)
-arg\Gamma(\frac{1}{2}-\frac{1}{2}\sqrt{1+4\gamma_1^2+4\gamma_2^2}+\varepsilon+i\eta)+\frac{R\lambda}{a}\right]}=-1.
\end{eqnarray}
Therefore, the corrected energy eigenvalue equation in a closed form becomes,
\begin{eqnarray}\label{eigen}
arg\Gamma(2i\lambda)-arg\Gamma(\frac{1}{2}+\frac{1}{2}\sqrt{1+4\gamma_1^2+4\gamma_2^2}+\varepsilon+i\eta)\\
-arg\Gamma(\frac{1}{2}-\frac{1}{2}\sqrt{1+4\gamma_1^2+4\gamma_2^2}+\varepsilon+i\eta)+\frac{R\lambda}{a}&=&(n_r+\frac{1}{2})\pi,
\,\, n_r=0,\pm1,\pm2,...,\nonumber
\end{eqnarray}
where $n_r$ is the radial node number. The quantum number is
n$=$n$_{r}+1$. In order to test the accuracy of Eq.(\ref{eigen}), we
calculate the single particle energy levels of neutron rotating
around $^{56}$Fe nucleus by using the potential parameters given in
Ref.\cite{berkdemir}. The Woods-Saxon potential parameters are
$V_{0}$=$40.5+0.13A$=$47.78$ MeV, R$=$4.9162 fm and a$=$0.6 fm. Here
$A$ is the atomic mass number of $^{56}$Fe nucleus. The reduced mass
consists of neutron mass $m_n=1.00866u$, and $^{56}$Fe core mass
$m_A=56u$. In Table \ref{Table}, we show agreement between the
energy eigenvalue equation given by Eq.(\ref{eigen}) and the
numerical calculation obtained by Gamow code\cite{gamow} for the
neutron plus $^{56}$Fe nucleus system with several $n_r$ quantum
numbers and $W_0$ parameters. There are small inaccuracy in the
analytic and numeric results since we have made the approximation in
Eq.(\ref{eigen2}). It might be seen that neutron is unbound for
$n_r=3,W_0=0$ and Eq.(\ref{eigen}) gives $E_{n_r\ell}=62.9775$MeV,
but this finding is not acceptable as physically and is only a
mathematical result which satisfies Eq.(\ref{eigen}). There are
similar situations for $n_r=3, W_0=50$MeV; $n_r=3, W_0=100$MeV; and
$n_r=2,3, W_0=-50$MeV. However there are the quasi-bound states
(resonant states) for $n_r=2, W_0=-100$ MeV since the nuclear
potential has a potential barrier inducing the resonant states in
Fig.\ref{potplot}. Our result for $n_r=2, W_0=-100$ MeV is
physically incorrect. In order to calculate the energy eigenvalues
of the resonant states, the Complex Scaling Method(CSM) can be
used\cite{cms}. Another interesting point is that we cannot
calculate the bound state energy eigenvalues by using
Eq.(\ref{eigen}) for $n_r=2,3, W_0=100$MeV since the right and left
sides of Eq.(\ref{eigen}) do not intersect in real line. It should
be noted that the $\ell$-state solution of the generalized
Woods-Saxon potential in terms of the Pekeris approximation is valid
only for small $\alpha$ values.

\begin{figure}
\includegraphics[width=\textwidth]{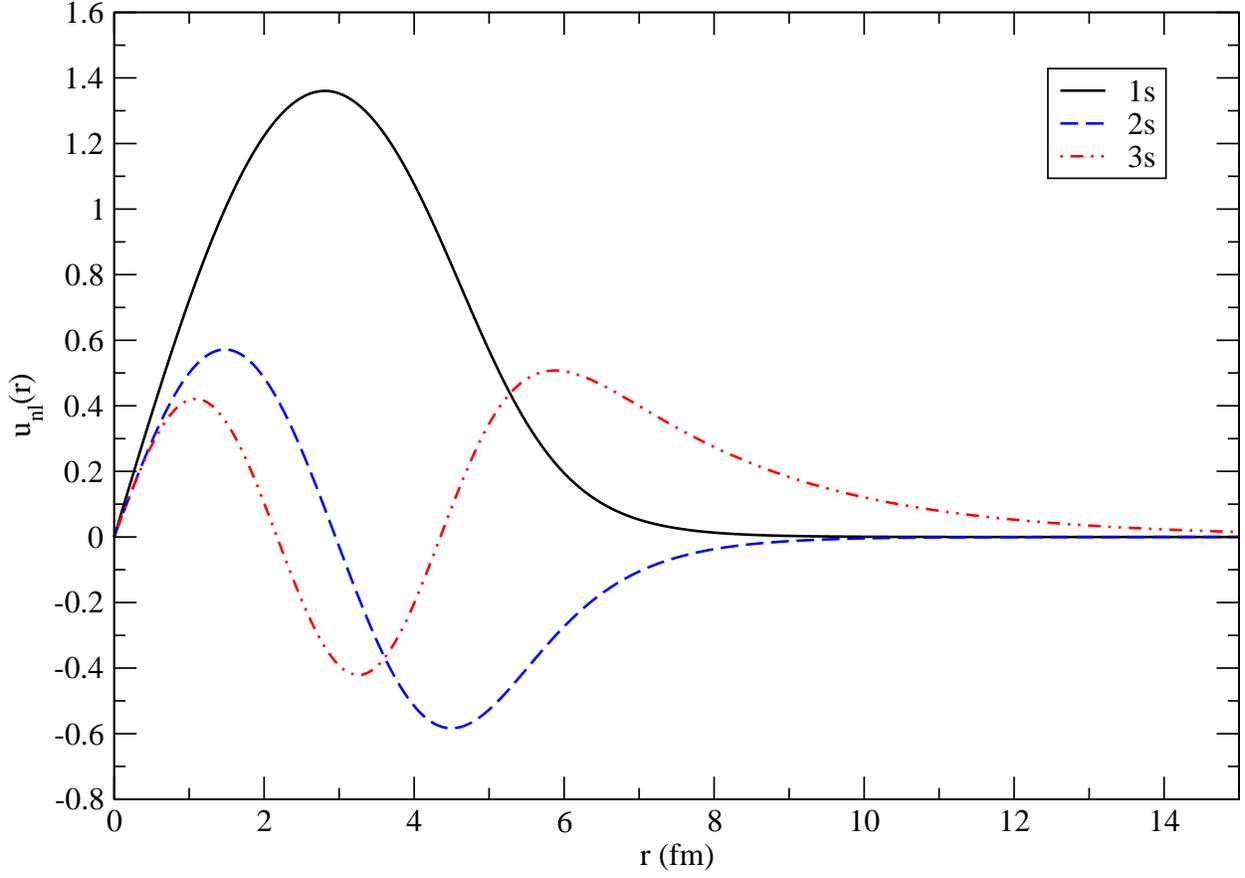}
\caption{The unnormalized wave function of the generalized
Woods-Saxon potential for $V_{0}$$=$40.5+0.13A MeV, $W_0$$=$50 MeV,
R$=$4.9162 fm and a$=$0.6 fm potential parameters with several $n$
quantum numbers.}\label{waveplot}
\end{figure}

The radial wave function corresponding to the eigenvalue equation
Eq.(\ref{eigen}) in terms of Eq.(\ref{ansatz}) and
Eq.(\ref{hyperabc}) can be written in a closed form as follows,
\begin{eqnarray}\label{radyalwave}
 u_{n\ell}(r)&=&N\left(\frac{1}{1+e^{\frac{r-R}{a}}}\right)^\varepsilon\left(1-\frac{1}{1+e^{\frac{r-R}{a}}}\right)^\eta  \\
&\times&{_{2}}F_{1}(\frac{1}{2}-\frac{1}{2}\sqrt{1+4\gamma_1^2+4\gamma_2^2}+\varepsilon+\eta,\frac{1}{2}+\frac{1}{2}\sqrt{1+4\gamma_1^2+4\gamma_2^2}+\varepsilon+\eta;1+2\varepsilon;\frac{1}{1+e^{\frac{r-R}{a}}}),\nonumber
\end{eqnarray}
where $N$ is the normalization constant. The unnormalized wave
function fulfilling the boundary conditions is shown in
Fig.\ref{waveplot} for several radial quantum numbers $n$.

\section{Conclusion}\label{conclude}
We have studied the approximate analytical solution of the
Schr\"{o}dinger equation in the presence of the generalized
Woods-Saxon potential by using the Pekeris approximation and the
properties of the Hypergeometric functions for any $\ell$ states. We
have seen that the Nikiforov-Uvarov method cannot take into account
the correct boundary conditions for the generalized Woods-Saxon
potential. Therefore we have carefully examined the asymptotic
behavior of the wave function of the generalized Woods-Saxon
potential and have obtained the corrected eigenvalue equation and
the corresponding eigenfunction in the closed form for any $\ell$
states. We have also calculated the single particle energy level of
neutron rotating around $^{56}$Fe nucleus for the given potential
parameters in order to check the consistency between the analytical
and Gamow code results. We have shown that the obtained analytical
results in this study are in good agreement with the results
obtained by the Gamow code for $\ell$$=$0 state. The resonant state
solutions of the generalized Woods-Saxon potential in a closed form
are in progress. 
\section*{Acknowledgments}
Authors would like to thank T\"{U}B\.{I}TAK and Akdeniz University
for the their financial supports as well as Dr. A. Soylu for useful
comments on the manuscript. 

\newpage

\begin{table}[tbp]
\begin{center}
\begin{tabular}{ccccccccccccccc}\hline\hline
$W_{0}(MeV)$&& $n_r$ &&& $E_{n_r}^{Analytical}$(MeV)&&&
$E_{n_r}^{Gamow}$(MeV) \\\hline
 0     &&  0  &&&  -38.3004                    &&& -38.3002                  \\
 0     &&  1  &&&  -18.2254                    &&& -18.2227                   \\
 0     &&  2  &&&  -0.2678                     &&& -0.2663                     \\
 0     &&  3  &&&  62.9775                     &&& unbound                     \\
 50    &&  0  &&&  -41.1965                    &&& -41.1964                  \\
 50    &&  1  &&&  -23.8789                    &&& -23.8788                   \\
 50    &&  2  &&&  -3.6472                     &&& -3.6471                    \\
 50    &&  3  &&&  52.0232                     &&& unbound                    \\
 100   &&  0  &&&  -45.4453                    &&& -45.4446                  \\
 100   &&  1  &&&  -29.1659                    &&& -29.1642               \\
 100   &&  2  &&& undetermined                 &&& -7.8143                \\
 100   &&  3  &&& undetermined                 &&& unbound                \\
-50    &&  0  &&&  -36.2136                    &&& -36.2168                 \\
-50    &&  1  &&&  -12.8469                    &&& -12.8504                   \\
-50    &&  2  &&&   18.5701                    &&&  unbound                  \\
-50    &&  3  &&&   64.2816                    &&&  unbound                  \\
-100   &&  0  &&&  -34.5902                    &&& -34.5956                 \\
-100   &&  1  &&&  -8.0843                     &&& -8.0902                   \\
-100   &&  2  &&&  19.2098                     &&& 17.74+i(-8.36)
\\\hline\hline
\end{tabular}
\end{center}
\caption{Comparison of the analytical and numerical results for the
single particle energy levels of neutron orbiting around $^{56}$Fe
nucleus with several $W_{0}$ potential depth and $n_r$ quantum
numbers for $\ell=0$. The potential parameters are
$V_{0}=40.5+0.13A=47.78 $ MeV, $R=4.9162$ fm and $a=0.6$ fm. The
reduced mass consists of neutron mass $m_n=1.00866u$, and $^{56}$Fe
core mass which is $m_A=56u$ and its value is $\mu=0.990814u$. }
\label{Table}
\end{table}

\end{document}